\documentclass[aps,prd,english,superscriptaddress,11pt,notitlepage]{revtex4}
	\usepackage[colorlinks=true, pdfstartview=FitV,
linkcolor=blue, citecolor=blue, urlcolor=blue]{hyperref}

\usepackage{amsmath}
\usepackage{amssymb}
\usepackage{graphicx}
\usepackage{hhline}
\usepackage{textcomp}
\makeatletter
\usepackage{babel}
\newcommand{\bea}{\begin{eqnarray}}
\newcommand{\eea}{\end{eqnarray}}

\newcommand{\be}{\begin{equation}}
\newcommand{\ee}{\end{equation}}
\usepackage[active]{srcltx}

\begin{document}


 \title{Integral identities and universal relations for solitons}

 \author{Christoph Adam}
\affiliation{Departamento de F\'isica de Part\'iculas, Universidad de Santiago de Compostela and 
Instituto Galego de F\'isica de Altas Enerxias (IGFAE) Santiago de Compostela, E-15782, Spain}

\author{Alberto Garcia Martin-Caro}
\affiliation{
Department of Physics, University of the Basque Country UPV/EHU, Bilbao, Spain
}

\author{Carlos Naya}%

\affiliation{Universidad de Alcal\'a, Departamento de F\'isica y Matem\'aticas, 28805 Alcalá de Henares, Spain}

\affiliation{
 Institute of Theoretical Physics,  Jagiellonian University, Lojasiewicza 11, 30-348 Krak\'{o}w, Poland
}

\author{Andrzej Wereszczy\'{n}ski}

\affiliation{
 Institute of Theoretical Physics,  Jagiellonian University, Lojasiewicza 11, 30-348 Krak\'{o}w, Poland
}
 \affiliation{Department of Applied Mathematics, University of Salamanca, Casas del Parque 2, 37008 - Salamanca, Spain
}

 \affiliation{
 International Institute for Sustainability with Knotted Chiral Meta Matter (WPI-SKCM2), Hiroshima University, Higashi-Hiroshima, Hiroshima 739-8526, Japan
 }

\begin{abstract}
We show that any nonlinear field theory giving rise to static solutions with finite energy like, e.g., topological solitons, allows us to derive an infinite number of integral identities which any such solution has to obey. These integral identities can always be understood as being generated by field transformations and their related Noether currents. We also explain why all integral identities generated by coordinate transformations become trivial for Bogomolnyi-Prasad-Sommerfield (BPS) solitons, i.e., topological solitons which saturate a topological energy bound. Finally, we consider applications of these identities to a broad class of nonlinear scalar theories, including the Skyrme model. More concretely, 
we find nontrivial integral identities that can be seen as model-independent relations between certain physical properties of the solitons in such theories, and we comment on the possible connection between these new relations and those already found in the context of astrophysical compact objects.
We also demonstrate the usefulness of said identities to estimate the precision of the numerical calculation of soliton observables.

\end{abstract}
\maketitle



\maketitle

\section{Introduction}

Nonlinear field theories play a fundamental role in many areas of physics, reaching from fluid dynamics and condensed matter to nuclear and particle physics. One of the most distinguished features of such theories is the possible existence of solitons. 
Solitons are solutions with a localised energy density whose dispersion or decay is prevented precisely by the nonlinearity. Among field theories supporting solitons a particular role is played by topological solitons, whose absolute stability results from purely topological arguments. Topological solitons behave like particles in many respects, and their "particle number" can frequently be related to an integer topological charge $\mathcal{N}$ of the underlying field theory. Excellent accounts of topological solitons, their properties, their relevance and their applications can be found, e.g., in \cite{R,SM,Shnir}.

Owing to their nonlinearity, soliton solutions must be determined by numerical methods almost in all cases. These numerical calculations constitute a daunting task in many occasions, particularly in $d>1$ space dimensions, where a full $d$-dimensional minimization of the relevant energy functional or the solution of a system of partial differential equations (PDEs) is required. Exact, analytical identities which any solution must obey, therefore, constitute a valuable tool to gauge the quality and precision of the numerical methods used for a particular calculation.
It is the main purpose of the present letter to highlight the existence of an infinite number of integral identities for said theories and to provide some concrete examples for their applications.

In \cite{bjarke}, integral identities were already derived for a rather large class of nonlinear field theories supporting topological solitons. 
The construction of the integral identities in \cite{bjarke}, however, was restricted to effectively one-dimensional systems, where the Euler-Lagrange (EL) equations can be reduced to an ordinary differential equation (ODE) via symmetry reductions, usually assuming spherical symmetry.
This restriction to a certain degree limits the practical relevance of the results of \cite{bjarke} for the control of the numerical precision, because the true complexity of numerical calculations for soliton models in $d>1$ space dimensions usually only sets in when the full PDEs resulting from the unconstrained EL equations - or the related minimization problem for the full unconstrained energy functional  - have to be solved. In the present letter, we close this gap and derive an infinity of integral identities parametrized by arbitrary functions, which any solution of the corresponding soliton model has to obey. 

As a first application, we will discuss some particular examples of practical relevance for the semi-classical quantization of skyrmions in the Skyrme model, namely integral identities for the moments of inertia which show up in this semi-classical quantization procedure.
Our second application is based on the observation that the integral identity for the spin moment of inertia can, in fact, be re-interpreted as a relation between three apparently independent observables of the theory, namely the trace of the spin moment of inertia, the rms radius of the mass distribution, and the so-called D-term. We prove that this relation between the three observables is {\em universal} in the sense that it is not restricted to the Skyrme model but holds, in fact, for any Lorentz-invariant scalar field theory in 3+1 dimensional Minkowski space.

With respect to the class of models we consider,  we will be more restrictive than \cite{bjarke} in the present letter.
In \cite{bjarke}, both defects (vortices, monopoles, \ldots) and textures (lumps, skyrmions, hopfions, \ldots) were considered, but here we shall restrict to the case of textures, for concreteness.  
Our results, however, should be generalizable to the case of defects without difficulty. 

We shall find that the integral identities can always be understood as being generated by field transformations and their related Noether currents. Among these field transformations, there exists a certain subclass which are induced by coordinate transformations on physical space. For this subclass of coordinate transformations, the resulting integral identities together with some simple examples have already been considered in \cite{Manton}. 
Further, some integral identities of the type considered in the present paper have been introduced and applied in \cite{Hoyos1} under the name of "deformation constraints", using a slightly different approach.
Similar identities have also been extended to include the gravitational field in \cite{Herdeiro:2021teo,Herdeiro:2022ids}, both for self gravitating, smooth solitonic systems and hairy black holes. It would be interesting to extend the integral identities we find here to the case of self gravitating configurations, but we will leave such analysis for a forthcoming publication.

We use the mostly minus convention for the metric in Minkowski space, i.e., $A^\mu B_\mu = \eta^{\mu\nu} A_\mu B_\nu  = A_0 B_0 - \delta_{ij} A_i B_j$. In our concrete examples in 3+1 Minkowski space, $\mu = 0,1,2,3$ and $i=1,2,3$.
\section{The integral identities}

Concretely, we consider Poincare-invariant field theories depending on $N$ scalar fields
$\Phi^A$, $A=1 \ldots N$ in $d+1$-dimensional Minkowski space. 
Further, the lagrangian density $\mathcal{L} = \mathcal{L}(\Phi^A, \partial_\mu \Phi^A)$ is a sum of terms, $\mathcal{L} = \mathcal{L}_0 + \mathcal{L}_2 + \mathcal{L}_4 + \ldots = \sum_n \mathcal{L}_n$, where $\mathcal{L}_n$ is a homogeneous polynomial of degree $n$ in the field derivatives $\partial_\mu \Phi^A$. In particular, $\mathcal{L}_0 = -V(\Phi^A)$ is a potential term. In addition, it is frequently assumed that in $\mathcal{L}_n$ only such polynomials are allowed which lead to a lagrangian which is at most of second order in time derivatives. Together with Poincare invariance, this assumption is already quite restrictive, and only a few terms are allowed. 
We will assume this in the concrete examples we consider, although most of the results below do not depend on this assumption.

We want to find integral identities for static (topological soliton) solutions, therefore we restrict to static field configurations. For static fields the energy functional is $E= \int d^d x \mathcal{E}$ where $\mathcal{E} = - \mathcal{L} = \sum_n \mathcal{E}_n$.  To derive the integral identities and understand their relation with the Noether currents of the theory, we start from the static EL equations
\be \label{ELeq}
{\rm d}_i \frac{\partial \mathcal{E} }{\partial (\partial_i \Phi^A)} -
\frac{\partial \mathcal{E}}{\partial \Phi^A} =0
\ee
 where $\partial_j \equiv (\partial/\partial x_j)$ and ${\rm d}_j \equiv (d/d x_j)$ are the partial and total derivatives w.r.t. the cartesian spatial coordinates $x_j$, respectively. 
 
 The general idea now is to multiply the EL equation by an arbitrary field transformation 
 $\delta \Phi^A (x_i, \Phi^B ,\partial_k \Phi^C ,\ldots)$ and move   $\delta \Phi^A$ inside the total derivative, resulting in the generic expression ${\rm d}_i  \mathcal{J}_{{\rm N}, i} + \mathcal{H} =0$. Here, $\mathcal{J}_{{\rm N}, i}$ is the Noether current which corresponds to the transformation $\delta \Phi^A$, and $\mathcal{H}$ is a remainder which is zero only if  $\delta \Phi^A$ is a symmetry transformation of the energy functional. If we restrict to transformations $\delta \Phi^A$ such that the total derivative ${\rm d}_i  \mathcal{J}_{{\rm N}, i}$ integrates to zero, then we find the nontrivial integral identity $\int d^d x \, \mathcal{H} =0$ whenever $\mathcal{H}$ is nonzero.

 Up to now, we assumed that the scalar fields $\Phi^A$ are unconstrained and take values in $\mathbb{R}^N$ ("linear sigma models"). Many topological soliton models in $d>1$, however, require a different topology of the target space. In the simplest cases, they are of the non-linear sigma model type, with the scalar fields $\Phi^A$ taking values on the unit sphere $\mathcal{S}^{N-1}$, $\Phi^A \Phi^A =1$. Concretely, for $d=2$ and $N=3$ this leads to the baby Skyrme model, for $d=3$ and $N=4$ the Skyrme model and its generalizations, whereas $d=3 $ and $N=3$ leads to versions of the Skyrme-Faddeev model. The constraint $\Phi^A \Phi^A =1$ must be imposed by a Lagrange multiplier $\lambda (x)$, $\mathcal{E} \to \mathcal{E} + (\lambda/2) \Phi^A \Phi^A$, leading to the Euler-Lagrange equation
 $$
 {\rm d}_i \frac{\partial \mathcal{E} }{\partial (\partial_i \Phi^A)} -
\frac{\partial \mathcal{E}}{\partial \Phi^A} - \lambda \Phi^A=0.
 $$
 Multiplying by $\Phi^A$, using the constraint and resolving for $\lambda$, allows to re-express this equation as the original unconstrained EL equation multiplied by a projection operator perpendicular to $\Phi^A$, i.e., 
 \be
 \left( \delta^{AB} - \Phi^A \Phi^B \right) \left( {\rm d}_i \frac{\partial \mathcal{E} }{\partial (\partial_i \Phi^B)} - \frac{\partial \mathcal{E}}{\partial \Phi^B} \right) \equiv \Pi^{AB}
\left( {\rm d}_i \frac{\partial \mathcal{E} }{\partial (\partial_i \Phi^B)} - \frac{\partial \mathcal{E}}{\partial \Phi^B} \right) =0. \nonumber
 \ee
 If we now restrict the field transformations  $\delta \Phi^A (x_i, \Phi^B ,\partial_k \Phi^C ,\ldots)$ to transformations perpendicular to $\Phi^A$ such that $\delta \Phi^A \Pi^{AB} = \delta \Phi^B$, then the resulting local (non-)conservation equations ${\rm d}_i  \mathcal{J}_{{\rm N}, i} + \mathcal{H} =0$ are formally identical to the non-constrained case discussed above, i.e., the projection operator can be ignored. In the sequel, we shall always assume that $\delta \Phi^A \perp \Phi^A$ unless stated otherwise.

\subsection{Coordinate transformations}

 An infinitesimal coordinate transformation $x_j \to x_j + \delta x_j$ acts on $\Phi^A$ like 
 $\delta \Phi^A = \Phi^A_{,j}\delta x_j$.
 Multiplying \eqref{ELeq} by $\partial_j \Phi^A \equiv \Phi^A_{,j}$ in a first step and moving this factor inside the total derivative, we get
 $$
 {\rm d}_i \left( \Phi^A_{,j}  \frac{\partial \mathcal{E} }{\partial \Phi^A_{,i}} \right) - \Phi^A_{,ij} \frac{\partial \mathcal{E} }{\partial  \Phi^A_{,i}} - \Phi^A_{,j}\frac{\partial \mathcal{E}}{\partial \Phi^A} 
=0
$$
or
\be \label{EMtens}
{\rm d}_i T_{ij} \equiv {\rm d}_i \left( \Phi^A_{,j}  \frac{\partial \mathcal{E} }{\partial \Phi^A_{,i}} - \delta_{ij} \mathcal{E} \right) =0
\ee
where $T_{ij}$ are the space-space components of the energy-momentum tensor. We remark that for constrained fields $\Phi^A_{,j} \perp \Phi^A$, and our discussion is valid both for linear and nonlinear sigma models. 
Multiplying \eqref{EMtens} by $\delta x_j$ and moving this factor inside the total derivative, therefore, allows us to find both the corresponding Noether currents and their (non-)conservation equations.
Obviously, we might equally well use the conservation equation \eqref{EMtens} as a starting point, and derive the integral identities by multiplying eq. \eqref{EMtens} by different $\delta x_j$. This was the point of view taken in \cite{Manton}. $\delta x_j$ can, in principle, still depend on $x_i$ and on $\Phi^A$, but we shall restrict to proper coordinate transformations $\delta x_j(x_i)$ in the sequel.

In particular, for a translation $\delta x_j = a_j = \, $ const. we get
$$
{\rm d}_i \mathcal{J}_{T, i} \equiv {\rm d}_i   \left( a_j \Phi^A_{,j}  \frac{\partial \mathcal{E} }{\partial \Phi^A_{,i}} - a_i \mathcal{E} \right) =0,
$$
where $\mathcal{J}_{T, i}$ is the Noether current for a translation in the direction $a_i$. For a rotation 
we have to fix the dimension $d$. For $d=3$ an infinitesimal rotation is $\delta x_j = \epsilon_{jkl} \theta_k x_l$, and we get
$$
{\rm d}_i \left[ \epsilon_{jkl} \theta_k x_l \left( \Phi^A_{,j}  \frac{\partial \mathcal{E} }{\partial \Phi^A_{,i}} - \delta_{ij} \mathcal{E} \right) \right] = \epsilon_{jki} \theta_k \Phi^A_{,j} \frac{\partial \mathcal{E} }{\partial \Phi^A_{,i}}.
$$
If $\mathcal{E}$ depends on $\Phi^A_{,i}$ only via monomials which are invariant under rotations, like 
$\Phi^A_{,k}\Phi^A_{,k}$, $\Phi^A_{,k}\Phi^A_{,l}\Phi^B_{,k}\Phi^B_{,l}$, etc., then the r.h.s. is zero, leading to
\be
{\rm d}_i \mathcal{J}_{R, i} \equiv {\rm d}_i \left[ \epsilon_{jkl} \theta_k x_l \left( \Phi^A_{,j}  \frac{\partial \mathcal{E} }{\partial \Phi^A_{,i}} - \delta_{ij} \mathcal{E} \right) \right] =0
\ee
where $\mathcal{J}_{R, i}$ is the Noether current for a rotation about the infinitesimal angle $\theta = |\theta_i|$ and the axis $\theta_i/\theta$. This expression is also valid for $d=2$ if we assume $\theta_k = \theta \delta_{k3}$ and restrict all the other indices to the values $1,2$.

We will be particularly interested in dilatations $\delta x_j = \epsilon x_j$ for an infinitesimal $\epsilon$. As all expressions are exactly linear in $\epsilon$, we simply divide by this factor.  
Multiplying \eqref{EMtens} by $\delta x_j = x_j$ and moving this factor inside the total derivative, we get
$$
 {\rm d}_i   \left( x_j \Phi^A_{,j}  \frac{\partial \mathcal{E} }{\partial \Phi^A_{,i}} - x_i \mathcal{E} \right) - (\partial_j\Phi^A ) \frac{\partial \mathcal{E} }{\partial (\partial_j \Phi^A)}  + d\, \mathcal{E} =0.
$$
But $\mathcal{E} = \sum_n \mathcal{E}_n$ is a sum of homogeneous polynomials in $\partial_i \Phi^A$ which obey $(\partial_j\Phi^A ) \frac{\partial \mathcal{E}_n }{\partial (\partial_j \Phi^A)} = n \mathcal{E}_n$, by definition. We therefore find the 
 non-conservation equation for the Noether current 
$\mathcal{J}^i_{\rm D}$ of dilatations,
\be \label{bas-id}
{\bf{\rm d}}_i \mathcal{J}_{{\rm D},i} = \sum_n (n-d) \mathcal{E}_n  = \delta_{ij}T_{ij} \equiv T\, , \quad 
\mathcal{J}_{{\rm D},i} \equiv 
x_j (\partial_j \Phi^A) \frac{\partial \mathcal{E}}{\partial (\partial_i\Phi^A) } -x_i \mathcal{E} = \, T_{ij}x_j .
\ee
Integrating \eqref{bas-id} and assuming that the boundary term (the integral of the total derivative) vanishes, we get  the well-known virial (or Derrick) identity 
\be \label{virial}
 \int d^d x\, T =  \sum_n (n-d)E_n =0 \, , \quad  E_n \equiv \int d^d x \mathcal{E}_n. 
\ee
The assumption that the boundary term vanishes - which is true in all models we consider - implies that the Noether current vanishes (sufficiently fast),
\be
\int d^d x {\bf{\rm d}}_j \mathcal{J}_{{\rm D},j} =0 \quad \Rightarrow \quad \lim_{|\vec x| \to \infty} 
 \mathcal{J}_{{\rm D},j} =0
\ee
which is also implied by the condition of finite energy.

We can find more integral identities by multiplying eq. \eqref{bas-id} by arbitrary functions $f(\vec x, \Phi^A)$ and by moving this factor inside the total derivative.  In this letter, we will restrict to functions $f(x_i)$, leading to
\be
-f(\vec x) \, T -
(\partial_j f)  \, T_{ji}x_i + 
{\bf{\rm d}}_j \left( f(\vec x)  \, T_{ji}x_i \right) =0.  \label{gen-id}
\ee 
We assume that $f$ has no singularities and that its behaviour for large $|\vec x|$ is such that the total derivative term integrates to zero. Integrating \eqref{gen-id} then leads to
\be
\int d^d x \left( f(\vec x)  \, T +
(\partial_j f)  \, T_{ji}x_i \right) =0.  \label{int-id}
\ee
This equation represents an infinite family of integral identities, parametrized by an arbitrary function $f$, for all the models considered.
Some possible particular choices for $f$ are powers of $r=|\vec x|$ or, more generally, polynomials in the space coordinates $x^i$. The maximal allowed degree of these polynomials will depend on the model under consideration. In particular, for solitons with an exponentially decaying energy density, arbitrary powers are allowed.

 In the subsequent analysis we will especially focus on a case where $f(\vec x) = x_k x_l$. Then, 
 \be \label{full-D-id}
 \int d^d x\, \left( x_k x_l T + x_j x_k T_{jl} + x_j x_l T_{jk} \right) = 0
 \ee
 which, for the trace $f(\vec x) = \delta_{kl}x_k x_l =r^2$ reduces to
\be 
\int d^d x \left( r^2 \, T +
  \, 2 T_{ij}x_ix_j \right) =0. \label{rel-D}
  \ee

\subsection{General transformations}

Up to now, we restricted to coordinate transformations, i.e., field transformations of the type  
$\delta \Phi^A = \Phi^A_{,j}\delta x_j$. Obviously, the full set of transformation $\delta \Phi^A (x_i, \Phi^B ,\partial_k \Phi^C ,\ldots)$ which are not coordinate transformation is huge, and most of them will be of little practical relevance. We will, therefore, only consider the simplest geometric target space transformations in a first step, and then multiply by some particular functions later on, to arrive at the specific algebraic structures which are required for specific applications of the resulting integral identities.

The simplest transformation is a translation by a constant $\delta \Phi^A = \epsilon^A$. It is not perpendicular to $\Phi^A$ and, therefore, only valid for linear sigma models. In this case, choosing the $N$ translations $(\epsilon_B)^A = \epsilon \delta^{AB}$, multiplying \eqref{ELeq} by this translation and dividing by the constant $\epsilon$, we recover the original EL equation \eqref{ELeq}. Further, $\mathcal{J}_{{\rm FT},i}^B = \frac{\partial \mathcal{E} }{\partial (\partial_i \Phi^B)} $ is the Noether current for a constant field translation in the $B$ direction in field space (FT = field translation), which is conserved if $\mathcal{E}$ only depends on field derivatives, but not on the field itself.

The next simplest transformation is $\delta \Phi^A = \epsilon \Phi^A$, but this transformation is the infinitesimal version of the scale transformation $\Phi^A \to e^\epsilon \Phi^A$ on target space, which is, again, not compatible with the constraint $\Phi^A \Phi^A =1$. The transformation $\delta \Phi^A = \Phi^A$ can be used for the generation of integral identities in field theories where the constraint 
$\Phi^A \Phi^A =1$ is not imposed. 

Another simple linear transformation, which is compatible with $\Phi^A \Phi^A =1$, is an infinitesimal rotation in the field variables,
\be
\delta \Phi^{A_1} = \epsilon^{A_1 A_2 \ldots A_N}\Phi^{A_2}.
\ee 
Multiplying eq. \eqref{ELeq} by this transformation, we find
\be 
\epsilon^{A_1 A_2 \ldots A_N} \left[{\rm d}_i \left( \Phi^{A_2} \frac{\partial \mathcal{E}}{\partial \Phi^{A_1}_{,i}} \right)
- \Phi^{A_2}_{,i} \frac{\partial \mathcal{E}}{\partial \Phi^{A_1}_{,i}} -
 \Phi^{A_2} \frac{\partial \mathcal{E}}{\partial \Phi^{A_1}}\right] =0.
\ee 
Frequently, $\mathcal{E}$ depends on field derivatives only via invariant expressions like
$\Phi^A_{,k}\Phi^A_{,k}$, $\Phi^A_{,k}\Phi^A_{,l}\Phi^B_{,k}\Phi^B_{,l}$, etc., then the second term is zero.
On the other hand, if $\mathcal{E}$ depends on $\Phi^A$, as well - e.g., via a potential term $\mathcal{E}_0 = V(\Phi^A)$ - then it cannot be invariant under all target space rotations, because the only invariant term, $\Phi^A \Phi^A =1$, is trivial. We get
\be \label{chi-tr}
{\rm d}_i \mathcal{J}_{{\rm ch},i}^{A_3 \ldots A_N} = \epsilon^{A_1 A_2 \ldots A_N}  \Phi^{A_2} \frac{\partial \mathcal{E}}{\partial \Phi^{A_1}} \, , \quad \mathcal{J}_{{\rm ch},i}^{A_3 \ldots A_N}  \equiv 
\epsilon^{A_1 A_2 \ldots A_N} \Phi^{A_2} \frac{\partial \mathcal{E}}{\partial \Phi^{A_1}_{,i}} 
\ee
where $\mathcal{J}_{{\rm ch},i}^{A_3 \ldots A_N} $ is the Noether current for target space rotations which we call "chiral transformations", because target space rotations can be identified with chiral transformations for the Skyrme model, from which we will take our explicit examples.

Now we proceed exactly as in the case of coordinate transformations. That is to say, we multiply eq. \eqref{chi-tr} by an arbitrary function $f(x_i,\Phi^A)$ and move this function inside the total derivative to arrive at the generic expression ${\rm d}_i \mathcal{J}_{{\rm N},i} + \mathcal{H} =0$. Then we integrate and arrive at the integral identity $\int d^d x \, \mathcal{H} =0$.

\section{An example: moments of inertia in the Skyrme model}

For a particular example of an application of these integral identities, we choose the Skyrme model \cite{Skyrme,NM-book} with $d=3$ and $N=4$. The Skyrme model is frequently expressed in terms of an SU(2) valued field $\, \mathcal{U}$ which is related to $\Phi^A$ via $\, \mathcal{U}= \Phi^4 \mathbb{I} + i \tau^a \Phi^a$ where $a=1,2,3$, $\tau^a$ are the Pauli matrices and $\mathbb{I}$ is the $2\times 2$ identity matrix. Certain properties of the Skyrme model, like its behavior under chiral transformations, are more transparent in this matrix notation. For numerical calculations, on the other hand, 
the vector notation $\Phi^A$ is more convenient, and also the expressions for the moments of inertia are simpler in this notation.

In the simplest case of the original Skyrme model, the energy density consists of three terms, 
$\mathcal{E} = \mathcal{E}_{024} \equiv \mathcal{E}_0 + \mathcal{E}_2 +  \mathcal{E}_4$ where
\be \label{Sk-mod}
\mathcal{E}_0 = V(\Phi^4)\, , \quad \mathcal{E}_2 = \Phi^A_{,i}\Phi^A_{,i} \, , \quad \mathcal{E}_4 =
\left( \Phi^A_{,i}\Phi^B_{,j} - \Phi^A_{,j}\Phi^B_{,i} \right) 
\left( \Phi^A_{,i}\Phi^B_{,j} - \Phi^A_{,j}\Phi^B_{,i} \right) .
\ee
We shall, however, also consider a generalized Skyrme model $\mathcal{E}_{0246} \equiv \mathcal{E}_0 + \mathcal{E}_2 + \mathcal{E}_4 + \mathcal{E}_6  $ \cite{sextic}, which includes a further term which contains six powers of field derivatives,
\be
  \mathcal{E}_{6}=\frac{1}{6}\left( \epsilon^{ABCD} \epsilon_{ijk} \Phi^A \Phi^B_{,i} \Phi^C_{,j} \Phi^D_{,k} \right)^2
   \ee
or, equivalently,
\be
  \mathcal{E}_{6}  =   \left( \Phi^A_{,i}  \Phi^A_{,i} \right)^3 - 3  \left( \Phi^A_{,i}  \Phi^A_{,i} \right)  \left( \Phi^B_{,j}  \Phi^B_{,k} \right)^2 +2  \left( \Phi^A_{,i}  \Phi^A_{,j} \right)  \left( \Phi^B_{,j}  \Phi^B_{,k} \right)  \left( \Phi^C_{,k}  \Phi^C_{,i} \right)  .
\ee

We will also need the expressions
\bea
\frac{\partial \mathcal{E}_2}{\partial (\partial_j\Phi^A) } &=& 2 \partial_j \Phi^A \label{var-e2} \\
\frac{\partial \mathcal{E}_4}{\partial (\partial_j\Phi^A) } &=& 8 \left( \partial_k \Phi^B \partial_k \Phi^B \partial_j \Phi^A  - \partial_k \Phi^A \partial_k \Phi^B \partial_j \Phi^B  \right)  \label{var-e4} \\
\frac{\partial \mathcal{E}_6}{\partial (\partial_j\Phi^A) } &=& 6 \left[ \Phi^A_{,j} \left( \Phi^B_{,k} \Phi^B_{,k}\right)^2 - \Phi^A_{,j} \left( \Phi^B_{,k} \Phi^B_{,l} \Phi^C_{,k} \Phi^C_{,l} \right) \right]
\nonumber \\
&& - \; 12 \left[ \left( \Phi^B_{,k} \Phi^B_{,k} \right) \left( \Phi^A_{,l} \Phi^C_{,l} \Phi^C_{,j} \right) - \Phi^A_{,k} \left( \Phi^B_{,j} \Phi^B_{,l} \Phi^C_{,k} \Phi^C_{,l}\right) \right] .
\eea

We now want to show that the integral identities of the last section can be used to derive nontrivial constraints for the moments of inertia of skyrmions (=topological solitons of the Skyrme model) in the rigid body approximation. In the rigid body approximation, a skyrmion is allowed to move slowly (non-relativistically) along symmetry directions in configuration space, but it is not allowed to deform. Concretely, these symmetries are translations, rotations and iso-rotations, where we will restrict our considerations to rotations and iso-rotations. 
That is to say, we consider skyrmions which are rotating slowly and with a constant angular velocity both in physical space and in field space. More concretely, if $\Phi^A(x_i)$ is a static skyrmion (soliton solution), we consider the configuration 
\be
\left(\delta^{AC} - \epsilon^{ABC4}\bar \theta^B\right) \Phi^C (x_i + \epsilon_{ijk} \theta_j x_k)  \simeq \Phi^A(x_i) + 
\delta_{\rm spin} \Phi^A + \delta_{\rm iso} \Phi^A
\ee
where
\be
\delta_{\rm spin} \Phi^A = \epsilon_{ijk} \Phi^A_{,i}\theta_j x_k 
\ee
and
\be \delta_{\rm iso} \Phi^A = -\epsilon^{ABC4} \bar \theta^B \Phi^C \quad \Rightarrow \quad 
 \delta_{\rm iso} \Phi^a = -\epsilon^{abc} \bar \theta^b \Phi^c \; , \quad \delta_{\rm iso} \Phi^4 =0
\ee
where $a = 1,2,3$, etc. Here we took into account that isospin rotations are those rotations in field space which do not rotate the field component $\Phi^4$. We chose a minus sign in the isospin rotation, because then a "hedgehog" configuration $\Phi^a = (x^a/r) \sin f(r)$, $\Phi^4 = \cos f(r)$, is invariant under a combined rotation and isorotation about the same angle and axis, as is usually assumed.
Further,  $\theta_j = t \omega_j$, $\bar \theta^a = t \beta^a$, and $\omega_j$ and $\beta^a$ are the small, constant angular velocities in physical space and isospin space, respectively. 

For the calculation of the corresponding moments of inertia we need the full, time dependent lagrangian density. Concretely, terms in the lagrangian density which do not contain time derivatives are invariant under these rotations, because the rotations are symmetries. On the other hand, time derivatives, which are zero for static fields, now will give nonzero contributions,
\bea
\frac{d}{dt} \left(\Phi^a + \delta \Phi^a \right) =  \dot{\delta \Phi^a} &=&  \Phi^a_{,i} \epsilon_{ijk}\omega_j x_k - \epsilon^{abc} \beta^b \Phi^c  \nonumber \\
\frac{d}{dt} \left(\Phi^4 + \delta \Phi^4 \right) =  \dot{\delta \Phi^4} &=&  \Phi^4_{,i} \epsilon_{ijk}\omega_j x_k.
\eea
If we restrict to a quadratic dependence on first time derivatives, then this results in a quadratic form in the angular velocities, from which the moments of inertia can be read off immediately.

Concretely, the term 
\be
\mathcal{L}_2 = \partial_\mu \Phi^A \partial^\mu \Phi^A = \dot{\Phi}^A \dot{\Phi}^A - 
\partial_i \Phi^A  \partial_i \Phi^A 
 \ee
 leads to
 \be
 \dot{\Phi}^A \dot{\Phi}^A  \equiv \frac{1}{2} u^{(2)}_{jm}\omega_j \omega_m - w^{(2)}_{jb}\omega_j \beta_b + \frac{1}{2}v^{(2)}_{ab} \beta_a \beta_b 
 \ee
 where $ u^{(2)}_{jm}$, $ v^{(2)}_{jm}$, $ w^{(2)}_{jm}$ are the spin/isospin/mixed moment of inertia densities,
 \bea
 u^{(2)}_{jm} &=& 2\epsilon_{ijk}\epsilon_{lmn} \partial_i\Phi^A \partial_l \Phi^A   x_k x_n \label{u2} \\
 w^{(2)}_{jb} &=& 2\epsilon_{ijk}\epsilon^{abc} \partial_i\Phi^a x_k \Phi^c \label{w2} \\
 v^{(2)}_{bd} &=& 2\epsilon^{abc}\epsilon^{ade} \Phi^c \Phi^e. \label{v2}
 \eea
 For the quartic term
 \bea
 \mathcal{L}_4 &=& -\left( \partial_\mu\Phi^A\partial_\nu \Phi^B - \partial_\nu\Phi^A \partial_\mu \Phi^B \right) \left( \partial^\mu\Phi^A\partial^\nu \Phi^B - \partial^\nu\Phi^A \partial^\mu \Phi^B \right) \nonumber
 \\
 &=& 4\left( \dot{\Phi}^A \dot{\Phi}^A \partial_i \Phi^B  \partial_i \Phi^B - 
 \dot{\Phi}^A \dot{\Phi}^B \partial_i \Phi^A  \partial_i \Phi^B - (\partial_i \Phi^B  \partial_i \Phi^B)^2
 + \partial_i \Phi^A  \partial_i \Phi^B \partial_j \Phi^A  \partial_j \Phi^B \right)
 \eea
the part quadratic in time derivatives gives
\be
4\left( \dot{\Phi}^A \dot{\Phi}^A \partial_i \Phi^B  \partial_i \Phi^B - 
 \dot{\Phi}^A \dot{\Phi}^B \partial_i \Phi^A  \partial_i \Phi^B \right) = \frac{1}{2} 
 u^{(4)}_{jm}\omega_j \omega_m - w^{(4)}_{jb}\omega_j \beta_b + \frac{1}{2}v^{(4)}_{ab} \beta_a \beta_b 
 \ee
 where
 \bea
u^{(4)}_{jm} &=& 8 \epsilon_{ijk}\epsilon_{lmn}  x_k x_n \left(  
 \partial_i \Phi^A \partial_l \Phi^A \partial_p \Phi^B \partial_p \Phi^B - 
 \partial_i \Phi^A \partial_p \Phi^A \partial_l \Phi^B \partial_p \Phi^B\right) \\
 w^{(4)}_{jb} &=&8 \epsilon_{ijk}\epsilon^{abc}  x_k \Phi^c \left( \partial_i\Phi^a \partial_l \Phi^B  \partial_l \Phi^B 
 - \partial_l\Phi^a \partial_i \Phi^B  \partial_l \Phi^B \right) \label{w4} \\
 v^{(4)}_{be} &=&8 \epsilon^{abc} \epsilon^{def}\Phi^c \Phi^f \left( \delta^{ad} \partial_l \Phi^B  \partial_l \Phi^B -
 \partial_l \Phi^a  \partial_l \Phi^d \right) .
 \eea
Finally, the sextic term
\be
  \mathcal{L}_{6}=-\frac{1}{6}\left( \epsilon^{ABCD} \epsilon^{\mu \nu \rho \sigma} \Phi^A \Phi^B_\nu \Phi^C_\rho \Phi^D_\sigma \right)^2
   \ee
or, equivalentely,
\be
  \mathcal{L}_{6}  =   \left( \Phi^A_\mu  \Phi^{A \mu} \right)^3 - 3  \left( \Phi^A_\mu  \Phi^{A \mu} \right)  \left( \Phi^B_\nu  \Phi^{B}_\rho \right)^2 +2  \left( \Phi^A_\mu  \Phi^{A \nu} \right)  \left( \Phi^B_\nu  \Phi^{B \rho} \right)  \left( \Phi^C_\rho  \Phi^{C \mu} \right) 
\ee
leads to the following quadratic part in time derivatives
%
 \be
  3( \dot{\Phi}^A \dot{\Phi}^A) \left[ \left( \Phi^B_{,i}  \Phi^{B }_{,i} \right)^2 -   \left( \Phi^B_{,i}  \Phi^{B}_{,j} \Phi^C_{,i}  \Phi^{C}_{,j}  \right) \right] - 
6  \dot{\Phi}^A \dot{\Phi}^B  \left[ \Phi^A_{,i}  \Phi^{B}_{,i} \left( \Phi^C_{,j}  \Phi^{C}_{,j} \right) -   \Phi^A_{,i}  \Phi^{B}_{,j} \left( \Phi^C_{,i}  \Phi^{C}_{,j} \right) \right].
 \ee
The resulting expressions for the MoI densities are rather cumbersome, therefore we will restrict to the spin MoI case,

  \bea
\frac{1}{2} u_{jm}^{(6)} &=& 3 \epsilon_{ijk} \epsilon_{lmn} x_k x_n \left[  \Phi^A_i \Phi^A_l \left( \left( \Phi^B_p  \Phi^{B}_p \right)^2  - \Phi^B_p  \Phi^{B}_q \Phi^C_p  \Phi^{C}_q \right) 
\right. \nonumber \\
    &&-\;  2 \left. 
   \Phi^A_i \Phi^B_l  \left( \Phi^C_p  \Phi^{C}_p \Phi^A_q  \Phi^{B}_q  - \Phi^C_p  \Phi^{C}_q \Phi^A_p  \Phi^{B}_q \right) \right] .
 \eea

 We will find that it is possible to derive integral identities both for the spin moment of inertia (MoI) $U_{jm} = \int d^3 x \, u_{jm}$ and for the mixed moment of inertia (MoI) $W_{jb} = \int d^3 x \, w_{jb}$. The reason for this fact is most easily understood by studying the contributions from the quadratic term $\mathcal{L}_2$, \eqref{u2}-\eqref{v2}. Indeed,
 \eqref{u2} is quadratic in field derivatives, whereas \eqref{w2} is linear. On the other hand, it follows from \eqref{EMtens} and \eqref{chi-tr}  that an integral identity which is derived from a proper coordinate transformation is at least quadratic in field derivatives, whereas an integral identity from a geometric target space transformation is at least linear. Consequently, we will find that integral identities for $U_{jm}$ can be derived from a Noether current related to a coordinate transformation, and integral identities for $W_{jb}$ can be derived from a Noether current related to a target space transformation. On the other hand, $v^{(2)}_{bd}$ does not contain any field derivatives and, therefore, integral identities cannot be derived for the related isospin moment of inertia $V_{bd} $ with the methods developed in this paper.
 
\subsection{Integral identity for the spin MoI}
 
  First, we want to derive an integral identity for the spin moment of inertia $U_{jm}$.
 For simplicity, we shall just consider the trace $u^{(n)} = \delta_{jm} u^{(n)}_{jm}$, leading to
 \be
 \frac{1}{2}u^{(2)} = \epsilon_{ijk}\epsilon_{ljn} \partial_i\Phi^A \partial_l \Phi^A 
  x_k x_n = r^2 \mathcal{E}_2 - x^i x^l \partial_i\Phi^A \partial_l \Phi^A ,
 \ee
 \bea
 \frac{1}{2}u^{(4)} &=& 4\epsilon_{ijk}\epsilon_{ljn} x_k x_n 
 \left(  
 \partial_i \Phi^A \partial_l \Phi^A \partial_m \Phi^B \partial_m \Phi^B - 
 \partial_i \Phi^A \partial_m \Phi^A \partial_l \Phi^B \partial_m \Phi^B\right) \nonumber \\
 &=& 2r^2 \mathcal{E}_4 - 4 x_i x_l \left(  
 \partial_i \Phi^A \partial_l \Phi^A \partial_m \Phi^B \partial_m \Phi^B - 
 \partial_i \Phi^A \partial_m \Phi^A \partial_l \Phi^B \partial_m \Phi^B\right) ,
 \eea
 and to 
 \bea
\frac{1}{2} u^{(6)} &=& 3 \left( \delta_{il}\delta_{kn} - \delta_{in}\delta_{kl} \right)  x_k x_n \left[  \Phi^A_i \Phi^A_l \left( \left( \Phi^B_p  \Phi^{B}_p \right)^2  - \Phi^B_p  \Phi^{B}_q \Phi^C_p  \Phi^{C}_q \right) 
\right. \nonumber \\
    &&-\;  2 \left. 
   \Phi^A_i \Phi^B_l  \left( \Phi^C_p  \Phi^{C}_p \Phi^A_q  \Phi^{B}_q  - \Phi^C_p  \Phi^{C}_q \Phi^A_p  \Phi^{B}_q \right) \right] \equiv 3r^2 \left[ \mathcal{E}_6 - \mathcal{F}_6 \right] ,
 \eea
where we introduce the useful terms 
\bea \label{F246}
&& \mathcal{F}_2 = \Phi^A_{,r} \Phi^A_{,r} \, , \quad  \mathcal{F}_4 = \Phi^A_{,r} \Phi^A_{,r}
\Phi^B_{,j} \Phi^B_{,j} - \Phi^A_{,r} \Phi^B_{,r}\Phi^A_{,j} \Phi^B_{,j} , \nonumber \\
&& \mathcal{F}_6 = \Phi^A_{,r}\Phi^A_{,r} \left[ \left(\Phi^B_{,p}\Phi^B_{,p} \right)^2 - \Phi^B_{,p} \Phi^B_{,q} \Phi^C_{,p} \Phi^C_{,q} \right] 
-\; 2\Phi^A_{,r} \Phi^B_{,r} \left[ \Phi^C_{,p}\Phi^C_{,p} \Phi^A_{,q} \Phi^B_{,q} - 
\Phi^C_{,p}\Phi^C_{,q} \Phi^A_{,p} \Phi^B_{,q} \right] ,
\eea
and we used that $x_j \partial_j \Phi^A = r\partial_r \Phi^A \equiv r \Phi^A_{,r}$.

 Finally, we get
 \be
 u = u^{(2)} + u^{(4)} + u^{(6)}  = 2r^2 \left( \mathcal{E}_2 +2 \mathcal{E}_4 + 3 \mathcal{E}_6 - \mathcal{F}_2 - 4 \mathcal{F}_4 -3 \mathcal{F}_6 \right)
 \ee

 For the trace of the moment of inertia tensor $U = \delta_{ij} U_{ij}$ this leads to
 \be 
 U = 2\int d^3 x \, r^2 \left( \mathcal{E}_2 +2 \mathcal{E}_4 +3 \mathcal{E}_6 - \mathcal{F}_2 - 4 \mathcal{F}_4 -3 \mathcal{F}_6 \right) .
 \ee
 
Now we want to derive an integral identity for $U$. For this purpose, we use the identity \eqref{bas-id} which is based on the Noether current of dilatations. 
We multiply \eqref{bas-id} by $x_k x_l$ and move this factor inside the total derivative, resulting in
\bea
 x_k x_l \left( (d+2)\mathcal{E} - \sum_n n \mathcal{E}_n\right) - x_i (\partial_i \Phi^A ) \left( x_k \frac{\partial \mathcal{E}}{\partial (\partial_l\Phi^A) } + x_l \frac{\partial \mathcal{E}}{\partial (\partial_k\Phi^A) } \right) + && \nonumber \\
+\, {\bf{\rm d}}_i \left( x_k x_l \left( x_j (\partial_j \Phi^A ) \frac{\partial \mathcal{E}}{\partial (\partial_i\Phi^A)}
- x_i \mathcal{E} \right) \right) &=& 0.
\eea
For the application we have in mind the contraction with $\delta_{lk}$ is sufficient, leading to
\bea
 r^2 \left( (d+2)\mathcal{E} - \sum_n n \mathcal{E}_n\right) - 2 x_i (\partial_i \Phi^A ) \left( x_k \frac{\partial \mathcal{E}}{\partial (\partial_k\Phi^A) }  \right) + && \nonumber \\
+\, {\bf{\rm d}}_i \left( r^2 \left( x_j (\partial_j \Phi^A ) \frac{\partial \mathcal{E}}{\partial (\partial_i\Phi^A)}
- x_i \mathcal{E} \right) \right) &=& 0. \label{INT-1}
\eea
For the energy density of the Skyrme model we find
\bea
 r^2 \left( 5 \mathcal{E}_0 + 3 \mathcal{E}_2 + \mathcal{E}_4 - \mathcal{E}_6 \right) - 2 x_i x_k 
  (\partial_i \Phi^A ) \left( \frac{\partial \mathcal{E}_2}{\partial (\partial_k\Phi^A) }  
  + \frac{\partial \mathcal{E}_4}{\partial (\partial_k\Phi^A) }
  + \frac{\partial \mathcal{E}_6}{\partial (\partial_k\Phi^A) } 
  \right)
&=& 0
\eea
or
\bea
 r^2 \left( 5 \mathcal{E}_0 + 3 \mathcal{E}_2 + \mathcal{E}_4 - \mathcal{E}_6 - 4 \mathcal{F}_2 - 16 \mathcal{F}_4 - 12 \mathcal{F}_6 \right) +\, {\bf{\rm d}}_i \left[ \cdots \right] &=& 0
 \eea
where $\mathcal{F}_2$, $\mathcal{F}_4$ and $\mathcal{F}_6$ are defined in \eqref{F246}.
Integrating and assuming that the boundary term vanishes, we find the integral identity 
\be \label{mom-ii}
\int d^3 x \, r^2 \left( 5 \mathcal{E}_0 + 3 \mathcal{E}_2 + \mathcal{E}_4 - \mathcal{E}_6 \right) = 4 \int d^3 x\, r^2 \left( 
\mathcal{F}_2 + 4 \mathcal{F}_4 + 3 \mathcal{F}_6 \right) .
\ee
This allows us to rewrite $U$ like
 \be \label{mom-ii-2}
 U 
 = \int d^3 x \, r^2 \left( \frac{1}{2}\mathcal{E}_2 + \frac{7}{2} \mathcal{E}_4 + \frac{13}{2} \mathcal{E}_6 - \frac{5}{2} \mathcal{E}_0 \right) .
 \ee
 This result can be used in two ways. Either, it provides us with a simplified expression for the spin moment of inertia $U$. Or it constitutes a nontrivial identity for the trace of the moment of inertia tensor, which can be used to control the precision of our numerical calculations. In the appendix, we provide some examples for this check of the numerical precision for the standard massive Skyrme model \eqref{Sk-mod}, with quite good results.
 
In some cases, the skyrmion whose moment of inertia we want to calculate has sufficient symmetry such that its moment of inertia tensor is proportional to the identity, $U_{ij} = (U /3) \delta_{ij}$. In these cases, \eqref{mom-ii-2} is an integral identity for the full moment of inertia tensor.

 \subsection{Integral identity for the mixed MoI}

 Here, we shall restrict to the standard Skyrme model \eqref{Sk-mod} for simplicity.
 In a first step, we rewrite $w^{(n)}_{jb}$ in \eqref{w2}, \eqref{w4} with the help of the identity 
 $\epsilon_{ijk}\epsilon^{abc} = \delta_i^a \delta_j^b\delta_k^c \, \pm \, {\rm perm}$, leading to
 \be
 w_{ab} = w^{(2)}_{ab} + w^{(4)}_{ab} = \delta_{ab} \left( S^{(1)} - S^{(2)} \right) -T^{(1)}_{ab} - T^{(2)}_{ab} +T^{(3)}_{ab} + T^{(4)}_{ab}
 \ee
 where
 \bea
 S^{(1)} &=& 2 x_c\Phi^c \left( \Phi^d_{,d} + 4 \left( \Phi^d_{,d}\Phi^B_{,k} \Phi^B_{,k} - \Phi^B_{,d}\Phi^B_{,k}\Phi^d_{,k}\right)  \right) \\
 S^{(2)} &=& 2 x_c \Phi^d \left( \Phi^c_{,d} + 4 \left( \Phi^c_{,d} \Phi^B_{,k} \Phi^B_{,k} - \Phi^B_{,c}\Phi^B_{,k}\Phi^d_{,k}\right) \right) \\
 T^{(1)}_{ab}  &=& 2 x_c\Phi^c \left( \Phi^a_{,b} + 4 \left( \Phi^a_{,b} \Phi^B_{,k} \Phi^B_{,k} - \Phi^B_{,b}\Phi^B_{,k}\Phi^a_{,k} \right) \right) \label{t1} \\
 T^{(2)}_{ab} &=& 2 x_b\Phi^a \left( \Phi^c_{,c} + 4 \left( \Phi^c_{,c} \Phi^B_{,k} \Phi^B_{,k} - \Phi^B_{,c}\Phi^B_{,k}\Phi^c_{,k} \right) \right) \\
 T^{(3)}_{ab} &=& 2 x_c\Phi^a \left( \Phi^c_{,b} + 4 \left( \Phi^c_{,b} \Phi^B_{,k} \Phi^B_{,k} - \Phi^B_{,b}\Phi^B_{,k}\Phi^c_{,k} \right) \right) \\
 T^{(4)}_{ab} &=& 2 x_b\Phi^c \left( \Phi^a_{,c} + 4 \left( \Phi^a_{,c} \Phi^B_{,k} \Phi^B_{,k} - \Phi^B_{,c}\Phi^B_{,k}\Phi^a_{,k} \right) \right) . \label{t4}
 \eea
 
 For the integral identity, we now use the identity for the target space rotations \eqref{chi-tr}, adapted to the case of isopin rotations in the Skyrme model,
 \be
 {\rm d}_i \left( \epsilon^{cde} \Phi^d \frac{\partial \mathcal{E}}{\partial \Phi^c_{,i}} \right) =0.
 \ee
 Multiplying by $\epsilon^{fae}x_f x_b$, evaluating the two $\epsilon$ symbols and moving the factor inside the total derivative, we arrive at
\be
 {\rm d}_i \left( x_c x_b \Phi^a \frac{\partial \mathcal{E}}{\partial \Phi^c_{,i}} - x_c x_b \Phi^c \frac{\partial \mathcal{E}}{\partial \Phi^a_{,i}} \right) - x_b \Phi^a \frac{\partial \mathcal{E}}{\partial \Phi^c_{,c}} - x_c \Phi^a \frac{\partial \mathcal{E}}{\partial \Phi^c_{,b}} + x_b \Phi^c \frac{\partial \mathcal{E}}{\partial \Phi^a_{,c}}
 + x_c \Phi^c \frac{\partial \mathcal{E}}{\partial \Phi^a_{,b}} =0.
 \ee
 We remark that, by construction, this identity can provide an integral identity only for the traceless part of the mixed MoI tensor.
 
 Finally, inserting the Skyrme model energy density $\mathcal{E} = \mathcal{E}_0 + \mathcal{E}_2 + 
 \mathcal{E}_4$ and using \eqref{var-e2} and \eqref{var-e4}, we arrive at
 \be
  {\rm d}_i [\ldots ]  +  \left( T^{(1)}_{ab} - T^{(2)}_{ab} - T^{(3)}_{ab} + T^{(4)}_{ab} \right) =0,
  \ee
  leading to the integral identity 
  \be \label{mixed_Ident}
  \int d^3 x \,  \left( T^{(1)}_{ab} - T^{(2)}_{ab} - T^{(3)}_{ab} + T^{(4)}_{ab} \right) =0
  \ee
  for the four tensor contributions defined in \eqref{t1} - \eqref{t4}. Again, we have checked these integral identities against some numerical solutions of the model, see Appendix \ref{Appendix} for details.
 
\section{Universal relations}

\subsection{The Skyrme model example reconsidered}

In a first instant, the integral identity \eqref{mom-ii-2} derived for the spin moment of inertia of the Skyrme model appears to be a coincidental mathematical relation between some integrals defined within this specific model.  In this section, however, we will show that {\em i)} this identity can be interpreted as a relation between several relevant observables of the theory and {\em ii)} it can be rewritten in a much more general form that applies to a much broader class of non-linear theories.

First of all, let us notice that the spin MoI calculated in the previous section results from a particular approximation, namely from restricting the fluctuations about a skyrmion to a finite-dimensional subspace. Concretely, we considered the degrees of freedom of a rigid rotor (the symmetries of the energy functional), but some low-lying vibrational modes can, in principle be added. In the Skyrme model approach to nucleons and nuclei, these zero and low-frequency modes are then usually quantized, and the whole procedure is known under the name of "semiclassical quantization" or "semiclassical approach". The resulting spin MoI $U_{ij}$ should, therefore, be called the semiclassical spin MoI and must be distinguished from the mechanical spin MoI \footnote{It is interesting to observe that the semiclassical MoI was never called "moment of inertia" in \cite{ANW}. This name was exclusively reserved for the mechanical MoI in that paper.}
 \be \label{mec-MoI}
 \Lambda_{ij} = \int d^3 x (r^2 \delta_{ij} - x_i x_j)\mathcal{E}.
 \ee
This quantity can be also viewed as a sort of quadrupole moment. Indeed it is a linear combination of the usual traceless quadrupole moment
 \be 
Q_{ij} = \int d^3 x (r^2 \delta_{ij} - 3x_i x_j)\mathcal{E}.
 \ee
 and its generalized version $\tilde{Q}_{ij}=\int d^3 x\, x_ix_j \mathcal{E}$.

Further, the trace of the mechanical MoI is related to the rms radius of the local energy distribution of the soliton, or its mass radius \cite{GarciaMartin-Caro:2023},
\begin{equation}
    \Lambda = 2M\langle r^2_{m}\rangle,
\end{equation}
where $\Lambda = \delta_{ij}\Lambda_{ij}$ and $M=E$ is the static energy (mass) of the skyrmion. 

For the Skyrme model the trace of the mechanical moment of inertia reads 
 \be 
 \Lambda = 2 \int d^3 x\, r^2(\mathcal{E}_0 + \mathcal{E}_2 + \mathcal{E}_4 + \mathcal{E}_6),
 \ee
and for the difference we get (remember $T=\delta_{ij}T_{ij}$)
 \be \label{UL-diff}
 \Lambda - U =   \frac{3}{2}\int d^3 x \, r^2 \left( 3 \mathcal{E}_0 +\mathcal{E}_2  -\mathcal{E}_4 - 3 \mathcal{E}_6 \right) =  -  \frac{3}{2} \int d^3 x \, r^2 T,
 \ee
where the right hand side is proportional to the second moment of the virial identity. 
We now want to argue that this implies that the difference \eqref{UL-diff} is typically positive, although there are exceptions. The difference will be positive if the skyrmion solution has a decay behavior for large $r$ such that $\mathcal{E}_n$ decays faster than $\mathcal{E}_m$ for $n>m$, because the additional factor $r^2$ in the integrand of \eqref{UL-diff} enhances the contribution of the large $r$ region. $\mathcal{E}_n$ contains higher powers of the Skyrme field and higher powers of derivatives than $\mathcal{E}_m$, so typically this will be the case.

 One exception is provided by the so-called BPS Skyrme model \cite{BPS-Sk}, that is, the submodel $\mathcal{E}_{06} \equiv \mathcal{E}_{0} + \mathcal{E}_{6}$,
 which is known to possess the property that all static finite energy soliton solutions are, in fact, BPS solutions, i.e., solutions of a simpler, first-order "BPS" equation which saturate the topological energy bound which can be defined in the model. This BPS equation is equivalent to the equation $\mathcal{E}_6 = \mathcal{E}_0$ and obviously implies that not only the second moment, but {\em all} moments of the virial identity are identically zero. The underlying reason is that all moments of the virial identity are integral identities generated by some coordinate transformations, and we shall prove in Section V that integral identities generated by coordinate transformations are trivially satisfied for BPS solutions.
 
For a better understanding of the physical implications of formula \eqref{UL-diff} it is crucial to introduce the so-called $D$-term
\cite{Polyakov1,Polyakov2}:
\be
\frac{D}{M}\equiv \frac{2}{5}  \int d^3 x \left( \frac{1}{3}r^2 \, T - T_{ij} x_ix_j \right). \label{virial_D}
\ee 
The physical interpretation of the $D$-term is that it is a fundamental property of a particle at rest, like its mass and spin, which measures the spatial distribution of its internal forces or stresses \cite{Polyakov1,Polyakov2}. The theoretical and experimental determination of the $D$-term, as well as other gravitational form factors,
of nucleons and nuclei is a hot topic of current research (see e.g. the recent review \cite{GFFreview} and references therein). Further, field theoretic models of extended particles, like the Skyrme model or holography based models, are perfectly suited for the study of the $D$-term of nucleons and nuclei, because they imply a canonical definition of the energy-momentum tensor \cite{grav,HoloDterm}.

Using the integral relation (\ref{rel-D}), we can rewrite it as \cite{GarciaMartin-Caro:2023}
\be
\frac{D}{M}= \frac{1}{3}  \int d^3 x \, r^2 T \label{TD}
\ee
and finally arrive at a very simple expression for \eqref{UL-diff},
\be \label{Id-U}
\Lambda - U = - \frac{9}{2} \frac{D}{M}.
\ee
We remark that $(-D)$ is expected to be non-negative for stability reasons \cite{Polyakov1,Polyakov2}, which again implies the non-negativity of the difference $\Lambda - U$. 

The main physical implication of \eqref{Id-U}, however, is that it connects several observables which are, at first glance, rather unrelated, namely the (trace of the) spin moment of inertia $U$, the mechanical moment of inertia, $\Lambda$ (or trace of the generalized quadrupolar moment), and the $D$-term. This formula might appear as an accident or a miracle which just happens to hold in the Skyrme model. We will prove below that it is, in fact, a universal, {\it model independent} relation valid for any relativistic scalar field theory. 
\subsection{Universal I-D-Q relation }

We begin with the trace of the spin moment of inertia tensor of the soliton, $U\equiv \delta_{ij}U_{ij}$.
The definition of this tensor just generalizes the case of the Skyrme model considered in the last section and goes as follows. Consider an infinitesimal, time-dependent rotation of the solitonic solution, such that 
\be \label{inf-rot}
\frac{d}{dt} \left(\Phi^A + \delta \Phi^A \right) =  \dot{\delta \Phi^A} =  \Phi^A_{,i} \epsilon_{ijk}\omega_j x_k ,\quad \omega_k =\dot{\theta}_k 
\ee
For any Lorentz-invariant Lagrangian density, terms which do not contain time derivatives are invariant under such transformations, because of rotational symmetry. On the other hand, time derivatives, which are zero for static fields, now will give nonzero contributions. These correspond to the kinetic energy term associated to time-dependent rotations. If we restrict to a quadratic dependence on first time derivatives, then this results in a quadratic form in the angular velocities, from which the moments of inertia can be read off immediately.
Explicitly, the general expression for the spin moment of inertia density of a soliton is
\be
u_{ij}=\frac{1}{2}\frac{\partial^2}{\partial\omega_i\partial\omega_j}\left(\frac{\partial^2\mathcal{L}}{\partial\dot{\Phi}^A\partial\dot{\Phi}^B}\delta \dot{\Phi}^A\delta \dot{\Phi}^B\right) = 
\frac{1}{2} \frac{\partial^2\mathcal{L}}{\partial\dot{\Phi}^A\partial\dot{\Phi}^B} \Phi^A_{,k} \Phi^B_{,l}x_m x_n \left( \epsilon_{kim}\epsilon_{ljn} + \epsilon_{kjm}\epsilon_{lin} \right).
 \ee
For rotations about static soliton solutions, this expression is time independent, but still seems to depend on the kinetic part of the Lagrangian. For scalar field theories which are {\em i)} Lorentz invariant and, {\em ii)} no more than quadratic in time derivatives, however, this is not the case and $u_{ij}$ can be expressed entirely in terms of the static energy functional. Lorentz invariance will be imposed by assuming that $\mathcal{L}$ depends on first field derivatives only via the term
\be X^{AB} \equiv \Phi^{A}_{,\mu} \Phi^{B,\mu} = \dot{\Phi}^A \dot{\Phi}^B - \Phi^A_{,k} \Phi^B_{,k},
\ee
i.e., $\mathcal{L}(X^{AB}, \Phi^A)$. Imposing the condition of being no more than quadratic in time derivatives directly on an arbitrary lagrangian is probably difficult. We will, therefore, instead project on this quadratic part by replacing $u_{ij} $ by $u_{ij} \vert_{\omega_i =0}$, which is sufficient for our purposes. Obviously, this projection is an identity for lagrangians which fulfill condition  {\em ii)}. With these assumptions we get 
\bea
\left.  \frac{\partial^2\mathcal{L}}{\partial \dot{\Phi}^A \partial \dot{\Phi}^B}\right|_{\omega_i =0}
&=& \left.  \frac{\partial^2\mathcal{L}}{\partial X^{CD} \partial X^{EF}} 
\frac{\partial X^{CD}}{\partial \dot{\Phi}^A} \frac{\partial X^{EF}}{\partial \dot{\Phi}^B} \right|_{\omega_i =0} + \left.
\frac{\partial\mathcal{L}}{\partial X^{CD} } 
\frac{\partial X^{CD}}{\partial \dot{\Phi}^A \partial \dot{\Phi}^B} \right|_{\omega_i =0} = \nonumber \\
 &=& 2 \left. \frac{\partial\mathcal{L}}{\partial X^{AB} } \right|_{\omega_i =0} = \; -
2 \left. \frac{\partial\mathcal{E}}{\partial X^{AB} } \right|_{\omega_i =0} 
\eea
because $\frac{\partial X^{CD}}{\partial \dot{\Phi}^A} $ is linear in $\omega_i$, and $\mathcal{L} = -\mathcal{E}$
for static configurations. $u_{ij}$ can now be conveniently expressed in terms of 
\be
\Delta_{ij}\equiv T_{ij}+\delta_{ij}\mathcal{E}=(\partial_j \Phi^A) \frac{\partial \mathcal{E}}{\partial (\partial_i\Phi^A) },
\ee
because $\Delta_{ij}$ satisfies the relation
\be
\Delta_{ij} = \left( \partial_j \Phi^A\right) \frac{\partial \mathcal{E}}{\partial (\partial_i\Phi^A) } = \Phi^A_{,j} \frac{\partial\mathcal{E}}{\partial X^{BC}} \frac{\partial X^{BC}}{\partial \Phi^A_{,i}} = -2 \Phi^A_{,j} \frac{\partial\mathcal{E}}{\partial X^{BC}} \delta^{AB} \Phi^C_{,i}
= - 2  \frac{\partial\mathcal{E}}{\partial X^{AB} } 
\Phi^A_{,i} \Phi^B_{,j} .
\ee
It follows that 
\be \label{full-uij} 
u_{ij}=
\frac{1}{2} \Delta_{kl}x_m x_n \left( \epsilon_{kim}\epsilon_{ljn} + \epsilon_{kjm}\epsilon_{lin} \right)
 \ee
and for the trace
\be \label{local-id-u-Delta}
u\equiv \delta_{ij}u_{ij} = \Delta_{ij} \left( \delta_{ij} r^2 - x_i x_j \right). 
\ee

Now we consider the trace of the spin MoI 
\be
  I\equiv  U=\int d^3x \, u= \int d^3x \Delta_{ij} \left( \delta_{ij} r^2-x_ix_j \right) = \int d^3x \left(T_{ij} +\delta_{ij}\mathcal{E} \right) \left( \delta_{ij} r^2-x_ix_j \right),
\ee
where we introduce the standard notation $I$ for the moment of inertia for convenience.
With the help of (\ref{rel-D}), this can be rewritten as
\be
2 \int d^3 x \, r^2\mathcal{E}  - U = -\int d^3x \,r^2 \left( T +\frac{1}{2}T \right)=- \frac{3}{2} \int d^3 x\, r^2 T.
\ee
Thus, finally we find
\be
\Lambda - U = - \frac{9}{2} \frac{D}{M}, 
\ee
which ends the proof. Introducing $Q\equiv \Lambda$ which underlines the relation of this quantity 
with the trace of the generalized quadrupolar moment, this can be re-expressed as
\be
Q - I = -\frac{9}{2} \frac{D}{M}, \label{IDQ}
\ee
constituting the IDQ relation announced in the title of this section.

With the help of eq. (\ref{full-uij}), the integral identity (\ref{full-D-id}) and some simple algebra,
a similar identity can be found for the full spin moment of inertia $U_{ij}$, 
\be
U_{ij} = \int d^3 x\, u_{ij} = \Lambda_{ij} + \int d^d x \, \left( \frac{3}{2} \delta_{ij} r^2 T - 2x_i x_j T - r^2 T_{ij} \right) .
\ee

The identity \eqref{IDQ} can be seen as a nontrivial virial identity, namely, an integral relation involving only different components of the stress-energy tensor when computed over solitonic solutions. However, it is reminiscent of some well-known ``universal'' (i.e. model-independent) relations in the context of compact, gravitating objects such as neutron stars \cite{ILQ} or boson stars \cite{Universalbosons}. Such relations typically involve the quadrupolar moment of the gravitational mass (in our case, $\Lambda \equiv Q$ corresponds to the quadrupole of the static energy), the moment of inertia (usually called I), and the so-called \emph{Love number}, which determines the reaction of the compact object against deformations due to stress produced by the spacetime curvature. In our relation \eqref{Id-U}, this role is played by the $D-$term, which can also be interpreted as a measure of the strength of the interaction of the soliton with a graviton \cite{grav}.
Therefore, the universal relation \eqref{IDQ} hints towards a possible origin of the universal relations for neutron stars in terms of some generalized virial relations involving the gravitational field.

  \subsection{Integral identity for Q-balls}
 
  We can extend our results for theories supporting $Q$-balls, that is {\it non-topological} solitons which are time-dependent, stationary solutions of the pertinent field equations \cite{FLS, C}. The corresponding Lagrange density reads
  \be
\mathcal{L}=\partial_\mu \Phi^* \partial^\mu \Phi - \mathcal{E}_0(|\Phi|),
  \ee
  where now the field $\Phi$ is a complex scalar. For a $Q$-ball, the time dependence enters only the phase factor,
  \be
  \Phi(x_i,t)=e^{i\omega t} \phi(x_i),
  \ee
 where $\omega$ is a frequency of the internal rotation and $\phi$ is the modulus of the complex scalar. This modulus can be treated as a new real field which is a solution of the {\it static} field equation obtained from an {\it effective} energy density
  \be
  \bar{\mathcal{E}}=\mathcal{E}_2+\bar{\mathcal{E}}_0 = (\partial_i \phi)^2 + \left(\mathcal{E}_0-\omega^2 \phi^2 \right).
  \ee
  Furthermore, the internal rotation leads to the appearance of the $U(1)$ Noether charge
  \be
  Q=2\omega \int d^3 x \phi^2.
  \ee
Now, we are ready to apply our formalism to this effective problem. We find the trace of the mechanical moment of inertia of the effective model
 \be
 \bar{\Lambda} = 2\int d^3 x r^2 \left( \mathcal{E}_2+\bar{\mathcal{E}}_0 \right)
 \ee
and  the trace of the spin moment of inertial of the effective model 
 \be
 \bar{U}=2 \int d^3 x r^2  \left( \mathcal{E}_2 - (\partial_r \phi)^2  \right).
 \ee
As before, these two quantities combine to a term proportional to the second moment of the trace of the spatial part of the energy-momentum tensor (of the effective theory)
\be
\bar{\Lambda}-\bar{U} = \frac{3}{2} \int d^3 x \, r^2 \left( \mathcal{E}_2+3\bar{\mathcal{E}_0}\right) \label{Q-ball-id1}
\ee
where we used  the identity (\ref{mom-ii}) for the $Q$-ball model
\be
\int d^3 x \, r^2 \left( 3\mathcal{E}_2+5\bar{\mathcal{E}}_0 \right) - \int d^3 x \, r^2 (\partial_r \phi)^2=0.
\ee
Now, we go back to the original theory. The true energy density of the time dependent $Q$-ball is 
\be
\mathcal{E}=\mathcal{E}_2+\mathcal{E}_0 + \omega^2 \phi^2 
\ee
and therefore the true mechanical moment of inertia (rms radius)
 \be
\Lambda = 2\int d^3 x r^2 \left( \mathcal{E}_2+\mathcal{E}_0 + \omega^2 \phi^2 \right) = \bar{\Lambda} + 4 \omega^2 \int d^3 x r^2  \phi^2 =\bar{\Lambda} + 2\omega Q^{(2)},
 \ee
  where 
\be
 Q^{(2)} = 2\omega \int d^3 x \, r^2 \phi^2.
\ee
is the second moment of the $U(1)$ charge density. 

Next, the spin moment of inertia does not explicitly depend on the potential. Hence, its trace remains unchanged $U=\bar{U}$ and we arrive at the following expression
\be
\Lambda - U = \frac{3}{2}  \int d^3 x r^2 \left(\mathcal{E}_2+3\mathcal{E}_0  \right)  - \frac{1}{2} \omega Q^{(2)} = - \frac{3}{2} \int d^3 x \, r^2 T  - \frac{1}{4} \omega Q^{(2)}.
\ee
Using the previously obtained relation between the trace of the spatial part of the energy-momentum tensor and the $D$ term (\ref{TD}) we find 
\be
\Lambda - U = - \frac{9}{2} \frac{D}{M} - \frac{1}{4} \omega Q^{(2)}.
\ee
Interestingly, the universal relation is modified by the second moment of the $U(1)$ charge density. Note that this new contribution is always negative. This should be contrasted with the term $-D/M$ which is always positive or equal zero. The non-negativity of $-D$ is, in fact, expected from stability arguments
\cite{Polyakov1,Polyakov2}. We also note that, in the generalization of Q-balls to the self gravitating case, also known as boson stars, the Noether charge is related to the total angular momentum of the solution \cite{Yoshida}. Therefore, the relation between the quadrupolar moment, the moment of inertia and the Love number will not be completely universal anymore, but depend on an additional parameter, namely, the angular momentum or spin of the solution. This is in fact what is found for several different models of boson stars in \cite{Universalbosons,Universalbosons2}.

 \section{BPS solitons}
 In general, topological soliton models allow us to derive a nontrivial lower bound, a so-called topological energy bound or BPS bound for the energy of static configurations, $E\ge E_{\rm BPS}$. Here, the bound energy $E_{\rm BPS}$ is a functional of the fields $\Phi^A$ which does not depend on the local properties of the fields but only on its boundary conditions. Typically, $E_{\rm BPS}$ can be expressed in terms of an integer topological index or "winding number" which can be defined in the model, $E_{\rm BPS} = C\cdot |\mathcal{N}|^\alpha$, where $\mathcal{N}$ is the topological index.  Further, $C$ is a fixed constant which only depends on the model but is the same for all field configurations. In many topological soliton models, 
 $E_{\rm BPS}$ is linear in $\mathcal{N}$, i.e., $\alpha =1$, but also field theories with a non-integer $\alpha$ are known. 
 
 The vacuum field configuration with topological index $\mathcal{N}=0$ always saturates the BPS bound,
 $E=E_{\rm BPS}=0$.
 One important question is whether the bound can be saturated in a given model for nontrivial field configurations with a nonzero $\mathcal{N}$, i.e., whether there exist BPS solutions with $E=E_{\rm BPS}$. It turns out that there exist so-called BPS models where all finite energy solutions saturate the bound. In some models there only exists a discrete set of BPS solutions (e.g.,  one solution with a fixed $\mathcal{N}$), up to symmetry transformations. 
 An example for this case is the version of the Skyrme model of Ref. \cite{Har}, which supports a BPS skyrmion in the $\mathcal{N}=1$ sector but not for $\mathcal{N}>1$. 
 And most topological soliton models do not support any nontrivial BPS solutions.
 
 We now want to explain briefly that all the integral identities \eqref{int-id} which are based on coordinate transformations are trivial for BPS solutions. The reason is that these identities involve the spatial energy-momentum tensor. But the spatial energy-momentum tensor is identically zero for BPS solutions. This fact is known for many concrete cases of BPS solutions, but usually not stated in  generality. A rather general proof was given in \cite{Schap}, but this proof made use of supersymmetry and, therefore, only holds for field theories which allow for supersymmetric extensions. This is true for many theories supporting BPS solutions, but not for all of them. For the Skyrme model of Ref. \cite{Har} mentioned above, for instance, a supersymmetric extension is not known and most likely does not exist. 
 
 Here we will give a rather general argument which covers all known cases of BPS solutions, both the cases covered by supersymmetry and BPS solutions for theories which do not allow for supersymmetric extensions. 
For our argument, it is useful to consider the energy functional $E$ depending on a general spatial metric $g_{ij}$, although we have $g_{ij} = \delta_{ij}$ for all models we consider. We use the notation  $g={\rm det} g_{ij}$ and now have to distinguish upper and lower indices. The energy functional $E[g_{ij}, \Phi^A]$ can be expressed like
\be
E[g_{ij}, \Phi^A] = \int d^d x \sqrt{g} \mathcal{E} (g_{ij}(x),\Phi^A(x) ,(\partial /\partial x^k) \Phi^B(x))
\ee
where $\mathcal{E}$ is a {\em scalar} function constructed from its arguments. 
Energy functionals of this type are invariant under coordinate transformations (diffeomorphisms) $x^i \to y^i = f^i (x)$, i.e.,
\be 
E[f^*(g_{ij}), f^*(\Phi^A) ] = E[g_{ij}, \Phi^A] \nonumber
\ee
where
\be 
 f^*(\Phi^A )  = \Phi^A (f(x)) \, , \quad f^*\left( g_{ij}\right) = \frac{\partial x^k}{\partial y^i}
 \frac{\partial x^l}{\partial y^j}g_{kl}(f(x)). \nonumber
 \ee
 But this implies that if we only transform $\Phi^A (x)$, i.e., replace $\Phi^A (x)$ by a different field
 $f^*( \Phi^A ) = \Phi^A (f(x))$ in the energy integrand, this transformation is equivalent to a transformation of the metric by the inverse of $f$,
 \be
 E[g_{ij}, f^*(\Phi^A) ] = E[(f^{-1})^*(g_{ij}), \Phi^A].
 \ee
 This argument was given, e.g., in \cite{Speight1}, where
 energy functionals of this type were called "geometrically natural".
 
 In a next step, we have to remember that the topological index $\mathcal{N}$ and the corresponding BPS energy are homotopy invariants, i.e., they are invariant under arbitrary smooth deformations of the fields. In particular, they are invariant under diffeomorphisms,
 \be
 E_{\rm BPS}[g_{ij}, f^* (\Phi^A) ] = E_{\rm BPS}[g_{ij}, \Phi^A ] \quad
 \Rightarrow \quad E_{\rm BPS}[(f^{-1})^*(g_{ij}), \Phi^A] = E_{\rm BPS}[g_{ij}, \Phi^A]
 \ee
 for arbitrary diffeomorphisms $f$. But this is possible only if $E_{\rm BPS}$ does not depend on the metric $g_{ij}$ at all, i.e., $E_{\rm BPS}=E_{\rm BPS}[ \Phi^A ]$. 

To continue, we now have to make some assumptions on the full energy functional for theories supporting BPS solutions,
\be 
E=\Delta E + E_{\rm BPS},
\ee
where $\Delta E \ge 0$ and $\Delta E =0$ for BPS solutions.
In most known cases of BPS models, the energy density is a sum of squares such that a completion of squares can be performed,
\be
\mathcal{E} = \sum_{i=1}^M (A_i^2 + B_i^2)= \sum_{i=1}^M (A_i -B_i)^2 + 2 \sum_{i=1}^M
A_i B_i \equiv \Delta \mathcal{E} + \mathcal{E}_{\rm BPS}.
\ee
Here, the $A_i$ and $B_i$ are functions of the fields and their first derivatives and, in addition,
\be
E_{\rm BPS} = \int d^d x \sqrt{g} \,\mathcal{E}_{\rm BPS} 
\ee
is a homotopy invariant. BPS solutions are then solutions to the set of first-order equations
\be
A_i - B_i =0 \, , \quad i = 1 ,\ldots ,M.
\ee
It is important to note that $\Delta \mathcal{E}$ is quadratic in the BPS operators
$A_i - B_i$. 

But it is not always possible to complete squares in the energy density such that the sum of the mixed terms leads to a homotopy invariant. This is the case, e.g., for the model of Ref. \cite{Har}. There, it is still possible to express the energy density as $\mathcal{E} = \Delta \mathcal{E} + \mathcal{E}_{\rm BPS}$ such that $\mathcal{E}_{\rm BPS}$ provides a homotopy invariant. The resulting $\Delta \mathcal{E}$ in this case, however, is a (rather complicated) polynomial in the BPS operators which contains quadratic, cubic and quartic terms. Importantly, a linear term is again absent.

These observations motivate the assumption which we make in the following for the energy density $\Delta \mathcal{E}$. Whenever the conditions for the saturation of the BPS bound can be expressed in terms of a set of first-order differential equations $A_i - B_i =0$, then $\Delta \mathcal{E}$ must be a polynomial in the corresponding BPS operators $A_i - B_i$ such that all terms in this polynomial are at least quadratic, i.e., there is no linear term. This condition is satisfied for all known BPS systems, to the best of our knowledge, although a general proof would clearly be desirable.  It is certainly satisfied for all theories which allow for supersymmetric extensions, because after the subtraction of the central extension of the SUSY algebra (which gives rise to $E_{\rm BPS}$), the remaining terms in the (bosonic part of the) energy density are at least quadratic in the supercharge densities which, on their part, are proportional to linear combinations of the BPS operators.

To close the argument, we now just need the fact that the (spatial) energy-momentum tensor can be expressed as a variation of the energy functional w.r.t. the spatial metric, i.e.,
 \be
\left. T^{ij}(x)\right| = 2g^{-1/2}\left. \frac{\delta E}{\delta g_{ij} (x)}\right| =
2g^{-1/2}\left. \left( \frac{\delta \Delta E}{\delta g_{ij} (x)} +
\frac{\delta E_{\rm BPS}}{\delta g_{ij} (x)}\right) \right| =0,
\ee
where the vertical line indicates that the expression is evaluated for a BPS solution. Here, the variation of $\Delta E$ is zero because, by our assumption, it is at least linear in the BPS operators as a consequence of the Leibniz rule, and BPS operators vanish for BPS solutions, by definition. 
The variation of $E_{\rm BPS}$ is zero because $E_{\rm BPS}$ does not depend on the metric.

On the other hand, there exist Noether currents like \eqref{chi-tr} which are related to transformations which act on the fields, i.e., which are not diffeomorphisms. These Noether currents still allow to find nontrivial integral identities also for BPS solutions, although we do not study this possibility in more detail in the present paper.

 \section{Discussion}




In this paper, we found infinitely many integral identities which any static solution of a topological soliton model has to satisfy, generalizing results from \cite{bjarke} and \cite{Manton}.
These identities imply rather severe restrictions for possible solutions, although they are, in general, not sufficient to completely determine them. In hindsight, the derivation of these identities might seem rather simple, but they imply, nevertheless, some rather nontrivial consequences and interesting applications. 

First of all, integral identities allow to estimate the precision of numerical calculations of soliton solutions and their related observables. The simplest (virial) identity has been frequently employed to estimate the precision of the soliton energies.  Another recent example where some extended virial constraints were used to gauge the precision of the numerical calculation of skyrmion crystals can be found in \cite{Leask1}.
The identities obtained in section III of the present paper specifically allow to gauge the precision in the calculation of the moments of inertia. The moments of inertia are the most relevant observables for the semi-classical quantization of skyrmions and their subsequent application to the description of physical nuclei. 
Further, the integrals defining these observables are more complicated than simple energy integrals and, therefore, specific integral identities are required to estimate their precision.

A second result is that these integral identities
lead to quite nontrivial relations between different relevant observables of the field theory under consideration. As a concrete example, we considered integral identities for the moments of inertia of the Skyrme model, but obviously there are many more possibilities. In fact, any moment of any Noether current of a theory can appear as an integrand of some integral identity. In our simple examples, we only used the Noether currents of dilatations and of target space rotations, but 
already the integral identities of the dilatation current and its moments allowed us to derive the universal relation \eqref{IDQ} which relates the mass rms radius squared (or the trace $Q\equiv 2 \delta_{ij} \tilde{Q}_{ij}$ of the generalized quadrupole moment $\tilde{Q}_{ij}$) with the trace of the moment of inertia $U$ (frequently called I) and the $D$-term. 
We want to emphasize the very general character of this universal relation (UR). While our more detailed discussion was restricted to scalar field theories, the UR is essentially derived from an integral identity which just relates different moments of the energy-momentum tensor. Closely related UR can therefore probably be found for any relativistic field theory which gives rise to an energy-momentum tensor.
Further, our universal relation is reminiscent of the well-known universal I-Love-Q relation for self-gravitating compact stars which up to now has been established only numerically, and may serve as a basis for a better theoretical understanding of the latter.

Obviously,
many more Noether currents like, e.g., additional linear combinations and higher moments of the  energy-momentum tensor, will give rise to integral identities which might lead to relations between relevant physical observables. 
In the particular case of the Skyrme model or other effective field theories of nuclear or hadron physics, the electroweak hadronic currents provide other relevant examples. 

A related issue is the calculation of form factors in nuclear and hadron physics. Here, the starting point is the matrix element of a hadronic current operator $\hat{\mathcal{J}}$ between an incoming ($i$) and an outgoing ($f$) hadronic or nuclear state, $\mathcal{J}_{if} (p,p') = \langle i, p| \hat{\mathcal{J}} |p' ,f \rangle $, where $p$ and $p'$ are the incoming and outgoing four-momenta. In the Breit frame, where there is no energy transfer between $i$ and $f$, and the spatial momenta are equal and opposite, $p = (p_0, \vec p)$, $p' = (p_0 , -\vec p)$, $\quad \Rightarrow \quad q \equiv p-p' = (0, \vec q) = (0 ,2\vec p)$, the hadronic matrix element only depends on the momentum transfer $\vec q$ and can be related to the corresponding matrix element in physical space by a simple Fourier transform,
\be
\mathcal{J}_{if} (\vec q) = \int d^3 x e^{i\vec x\cdot \vec q} \mathcal{J}_{if} (\vec x).
\ee
Now, if the exponent inside the integral is expanded in powers of $\vec q$, then the integrals multiplying these powers are precisely the higher moments of hadronic current matrix elements, which show up in the corresponding integral identities.
These identities will be particularly useful when it is easier to calculate the hadronic currents directly in physical space, $\mathcal{J}_{if} (\vec x)$, because then the first few terms in the expansion of the form factors are precisely the higher moments of the hadronic currents. This is the case, e.g., in effective field theories like the Skyrme model. Recent calculations of both gravitational and electroweak form factors in the Skyrme model can be found, e.g., in \cite{grav} and in \cite{Alb-thesis}, respectively.

The relation \eqref{Id-U} is also interesting from the experimental point of view, as it establishes a connection between physically observable magnitudes. Therefore, it would be interesting to check whether it is satisfied for simple nuclei. If it does, then it can serve as an indirect measurement method of the $D-$term, which is difficult to measure directly in scattering experiments.

There are many possibilities for generalizations to other field theories or other base space geometries.
One first possibility is the study of integral identities for models supporting defect type solitons like vortices or monopoles. Owing to their nontrivial behavior in the limit $|\vec x| \to \infty$, in the corresponding integral identities additional boundary terms ("boundary charges") will appear, as was already observed in \cite{bjarke}. Moreover, precisely due to the slow decay of such solutions with distance, vortices and monopoles can not rotate (indeed, it has been shown that no generalization of t'Hooft-Polyakov monopoles with finite angular momentum can exist in flat space \cite{Volkov}), which makes it difficult to define their moment of inertia, and hence the universal relation \eqref{Id-U} won't apply for such solutions.
Also solitons on non-flat spaces or fully coupled to gravity should allow for the derivation of similar integral identities.

Finally, we would like to comment on possible physical implications of the difference between the semi-classical spin MoI $U_{ij}$ and the mechanical spin MoI $\Lambda_{ij}$ in the Skyrme model discussed in Section IV.A. Both expressions are based on a rigid rotor approximation for the rotating skyrmion, which will be a good approximation for sufficiently low angular velocities if the model does not possess further space symmetries which would allow for deformations of the skyrmion at zero energetical cost. The difference between $\Lambda_{ij}$ and $U_{ij}$ is that the full mass density participates in the rigid rotation described by $\Lambda_{ij}$, whereas not all components of the skyrmion participate in the true rigid rotation described by $U_{ij}$. In other words, a general skyrmion solution could be interpreted as a two-component substance, where the normal matter component participates in the rotation and gives rise to the semiclassical spin MoI, whereas the second component behaves like a superfluid and does not rotate. More concretely, we found that this superfluid component is related to the second moment of the virial identity or, equivalently, to the $D$-term which, thus, could provide a measure for the amount of superfluidity present in a given skyrmion solution. This picture of skyrmions as composed of normal (a priori, solid or fluid) matter and a superfluid, obviously, has important implications for the application of the Skyrme model to nuclear matter. Interestingly, recently the concept of supersolid matter \cite{supersolid1, supersolid2}, that is, a type of matter composed of a normal solid component and a superfluid has been applied both to neutron stars \cite{supersolid3} and to finite nuclei \cite{supersolid4}, and our results could shed new light on these recent developments.
In any case, a detailed study of these implications will be provided elsewhere. Here we just want to mention that, in principle, it is known that the spin moments of inertia of nuclei are smaller than their mechanical MoI \eqref{mec-MoI}, and that this fact can be interpreted as the presence of a superfluid component \cite{Migdal}. This issue, however, has not yet been studied within the Skyrme model context, to the best of our knowledge.


\begin{acknowledgements} 

The authors acknowledge financial support from the Spanish Research State Agency under project PID2023-152762NB-I00, the Xunta de Galicia under the project ED431F 2023/10 and the CIGUS Network of Research Centres, the María de Maeztu grant CEX2023-001318-M funded by MICIU/AEI /10.13039/501100011033, and the European Union ERDF.
CN and AW are supported by the Polish National Science Center,
grant NCN 2020/39/B/ST2/01553. AGMC acknowledges financial support from the PID2021-123703NB-C21 grant funded by MCIN/ AEI/10.13039/501100011033/ and by ERDF, ``A way of making Europe''; and the Basque Government grant (IT-1628-22).
AW thanks Nick Manton for a very inspiring discussion about superfluidity in the Skyrme model.

\end{acknowledgements}

\appendix

\section{MoI integral identities within the massive Skyrme model}
\label{Appendix}

In this appendix, we want to apply the integral identities for the spin and mixed MoI to numerical solutions of the massive Skyrme model. The energy functional takes the form given in Eq. (\ref{Sk-mod}), where the potential is usually chosen to give mass to the pionic degrees of freedom behind the SU(2) formulation. In particular, it reads 
\be
V(\Phi^4) = 2 M_\pi^2 (1 - \Phi^4),
\ee
with $M_\pi$ a dimensional pion mass parameter which in the most common calibration of the model takes the value $M_\pi = 0.528$.

For this purpose, we found numerical configurations minimizing the static energy functional, i.e., Skyrmions, up to topological charge 4. We have used the method known as arrested Newton flow in a lattice of $100^3$ points and spacing $\Delta_x = 0.2$ (see e.g. \cite{Adam2024} for a brief explanation). Although these settings might be seen as to be lacking some accuracy, they will better support our idea of the integral identities as a measurement of numerical precision.

In table \ref{Num_Values} we have summarised the results concerning the integral identities of the spin MoI tensor as given by Eq. (\ref{mom-ii-2}). In the case of the mixed MoI identities, Eq (\ref{mixed_Ident}), all vanishing integrals are of the order $10^{-1}$ or lower. Furthermore, we have also included the check of the virial constraint in Eq. (\ref{mixed_Ident}), which for the massive Skyrme model reads, 
\be
E_4 -E_2-3 E_0 = 0.
\ee
Note that we have normalized it by the total energy of the Skyrmion to give a better idea of the precision of our solutions. Indeed, the values we got show us that even with the coarse lattice we considered the integral identities are fulfilled quite accurately, at least for the rather small topological charges we considered. We want to emphasize that the calculations shown here should be considered a mere proof of concept, and that a good precision should be expected for small topological charges. The full power of these integral identities will kick in when skyrmions of sufficiently large baryon number and their moments of inertia are calculated.

\begin{table} 
\begin{center}
\begin{tabular}{cccc}
\hline \hline
Topological charge \, & \, Virial \, & \, Spin MoI (numerical) \, & \, Spin MoI (identity) \\
\hline
\hline
1 & 0.004 & 185.4 & 186.2 \\
2 &  0.003  & 747.2 & 748.4 \\
3 & 0.002  & 1554.4 & 1557.0 \\
4 & 0.003  & 2578.4 & 2582.8 \\
\hline
\hline
\end{tabular}
\caption{Numerical values of the integral identities up to topological charge 4. In the second column we have the value of the virial constraint normalized by the Skyrmion energy. The third and fourth columns show the trace of the spin MoI tensor obtained by the numerical solutions and the integral identities, respectively.} \label{Num_Values}
\end{center}
\end{table}

\end{document}